\documentclass{article} \usepackage{slashed,feynmf,epsfig,amsmath,amssymb,enumitem}
\usepackage[papersize={8.5in,11in}]{geometry}  
\renewcommand{\vec}[1]{\boldsymbol{\mathrm{#1}}} 
\begin{document} \title{Variable Speed of Light Cosmology, Primordial
Fluctuations and Gravitational Waves} \author{J. W. Moffat\\~\\ Perimeter Institute for Theoretical Physics, Waterloo,
Ontario N2L 2Y5, Canada\\ and\\ Department of Physics and Astronomy, University of Waterloo, Waterloo,\\ Ontario N2L 3G1,
Canada} \maketitle




\begin{abstract}  A variable speed of light (VSL) cosmology is described in which the causal mechanism of generating primordial perturbations is achieved by varying the speed of light in a primordial epoch. This yields an alternative to inflation for explaining the formation of the cosmic microwave background (CMB) and the large scale structure (LSS) of the universe. The initial value horizon and flatness problems in cosmology are solved. The model predicts primordial scalar and tensor fluctuation spectral indices $n_s=0.96$ and $n_t=- 0.04$, respectively.  We make use of the $\delta{\cal N}$ formalism to identify signatures of primordial nonlinear fluctuations, and this allows the VSL model to be distinguished from inflationary models. In particular, we find that the parameter $f_{\rm NL}=5$ in the variable speed of light cosmology.  The value of the parameter $g_{\rm NL}$ evolves during the primordial era and shows a running behavior.

\end{abstract}

\maketitle

\section{Introduction}

Although inflationary models have been successful in fitting cosmological
data~\cite{Planck2015}, there have been issues raised about the fundamental consequences
of the models~\cite{Steinhardt}. In particular, the need for chaotic and eternal inflation models has raised the specter
of a multiverse cosmological scenario~\cite{Guth,Steinhardt2}. The question as to whether such a scenario can be falsified
by observations and the lack of predictability of inflation models has been a cause for concern. Moreover, the standard
single-field inflation models suffer from significant fine-tuning, such as the requirement of a slow-rolling potential and
the fine-tuning needed to fit the magnitude of the CMB amplitude~\cite{Guth2,Linde,Linde2,Hawking,Steinhardt3,Lyth}.

We shall consider a model based on the idea that the speed of light $c$ can have a significantly larger value in the very
early universe~\cite{Moffat,Moffat2,Magueijo,Barrow,Barrow2,Magueijo2,Ellis2,Farhoudi,Ellis,Shojaie,Magueijo3,Roshan}.
This assumption leads to a resolution of the horizon and flatness problems, thereby solving the initial value problem in
early universe cosmology. A bimetric variable speed of light (VSL) model~\cite{Moffat3,Moffat4,Moffat5,Moffat6,Magueijo4}
has been proposed with two metric tensors. One metric and its light cone describe a varying speed of light and a constant
speed of gravitational waves, while the other metric and its light cone describe a constant speed of light and a varying
speed of gravitational waves. This model cannot produce an observed value for relic gravitational waves and a non-zero
tensor mode spectral index $n_t$.

In the following, we will formulate a version of VSL based on an earlier VSL model with one metric. In Section 2, we
postulate a gravitational action in which the speed of light $c=c(x)$ is a dynamical field. The total action also contains
an action for a minimally coupled scalar ``seed'' field $\phi$, which will produce quantum primordial fluctuations. It
also contains an action inducing a spontaneous violation of Lorentz invariance~\cite{Moffat,Moffat2,Moffat7,Moffat8,Bluhm,Kostelecki}
by means of a non-vanishing vacuum expectation value of a vector field $\psi_\mu$. The Lorentz group $SO(3,1)$ is spontaneously broken resulting in SO(3,1)$\rightarrow O(3)\times R$, where $R$ is the absolute preferred time corresponding to the comoving time $t$ in the preferred frame
associated with the Friedmann-Lema\^{i}tre-Robertson-Walker (FLRW) metric. In Section 3, we develop the VSL cosmology and
we show how the model can solve the horizon and flatness problems in initial value cosmology. Section 4 presents the calculations of the power spectra and spectral indices for primordial scalar density perturbations and tensor gravitational waves. Section 5 we calculate the non-gaussianities in our VSL model. In Section 6 we end with concluding comments.

The VSL cosmology can remove the fine-tuning of the initial values in the standard big-bang cosmology and fit the
available observational data, such as an almost scale invariant, adiabatic and Gaussian scalar matter power spectrum and
the potential observation of a gravitational wave power spectrum.  As a viable alternative to standard inflationary
models, it can relieve the need for fine-tuning present in these models and not require a multiverse scenario in the form
of eternal inflation.

\section{The action and field equations}

We adopt the following action: 
\begin{equation} S=S_G+S_\psi+S_\phi+S_M, 
\end{equation} 
where 
\begin{equation}
S_G=\frac{1}{16\pi G}\int d^4x\sqrt{-g}\biggl[\Phi(R+2\Lambda)-\frac{\kappa}{\Phi}\partial^\sigma\Phi\partial_\sigma\Phi\biggr]. 
\end{equation} 
Here, $G$ is Newton's gravitational constant, $g={\rm det}(g_{\mu\nu})$, $R=g^{\mu\nu}R_{\mu\nu}$, $\Lambda$ is the cosmological constant, $\Phi(x)=c^4(x)$ and $\kappa$ is a dimensionless constant. The action $S_\psi$ is given by
\begin{equation} 
S_\psi=-\int d^4x\sqrt{-g}\biggl[\frac{1}{4}B^{\mu\nu}B_{\mu\nu}+W(\psi_\mu)\biggr], 
\end{equation} 
where $\psi_\mu$ is a vector field, $B_{\mu\nu}=\partial_\mu\psi_\nu-\partial_\nu\psi_\mu$, and $W(\psi_\mu)$ is a potential.
The scalar field action $S_\phi$ is defined by 
\begin{equation} 
S_\phi=\int d^4x\sqrt{-g}\biggl[\frac{1}{2}(\partial_\mu\phi\partial^\mu\phi)-V(\phi)\biggr], 
\end{equation} 
where $\phi$ is a scalar field and $V(\phi)$ is a potential. The matter action is $S_M=S_M(\phi_m, g_{\mu\nu})$ where $\phi_m$ denotes matter fields.

The energy-momentum tensor is 
\begin{equation} 
T_{\mu\nu}=T_{M\mu\nu}+T_{\psi\mu\nu}+T_{\phi\mu\nu}, 
\end{equation} 
where
\begin{equation} 
\frac{1}{\sqrt{-g}}\frac{\delta S_X}{\delta g^{\mu\nu}}=-\frac{1}{2}T_{X\mu\nu},\quad [X=M,\psi,\phi].
\end{equation} 
The $T_{\psi\mu\nu}$ and $T_{\phi\mu\nu}$ are given by 
\begin{equation}
T_{\psi\mu\nu}=B_{\mu\alpha}{B_{\nu}}^\alpha-\frac{1}{4}g_{\mu\nu}B^{\alpha\beta}B_{\alpha\beta} -2\frac{\partial
W(\psi_\mu)}{\partial g^{\mu\nu}}+g_{\mu\nu}W(\psi_\mu), 
\end{equation} and 
\begin{equation}
T_{\phi\mu\nu}=\partial_\mu\phi\partial_\nu\phi
-\frac{1}{2}g_{\mu\nu}g^{\alpha\beta}\partial_\alpha\phi\partial_\beta\phi -2\frac{\partial V(\phi)}{\partial g^{\mu\nu}}+g_{\mu\nu}V(\phi). 
\end{equation}

The variations of the action with respect to $g^{\mu\nu}$ and $\Phi$ yield the field equations: 
\begin{equation} 
\label{Graveqs} 
G_{\mu\nu}-g_{\mu\nu}\Lambda=\frac{8\pi G}{\Phi}T_{\mu\nu}+\frac{1}{\Phi}(\nabla_\mu\nabla_\nu
-g_{\mu\nu}\nabla^\alpha\nabla_\alpha)\Phi+\frac{\kappa}{\Phi^2}(\partial^\mu\Phi\partial_\mu\Phi-\frac{1}{2}g_{\mu\nu}\partial^\alpha\Phi\partial_\alpha\Phi),
\end{equation} 
\begin{equation}
\nabla^\alpha\nabla_\alpha\Phi=\frac{8\pi G}{3+2\kappa}T,
\end{equation}
where $G_{\mu\nu}=R_{\mu\nu}-\frac{1}{2}g_{\mu\nu}R$, $T=g^{\mu\nu}T_{\mu\nu}$ and $\nabla_\sigma$ is the covariant derivative with respect to the metric $g_{\mu\nu}$. We also obtain the field equations: 
\begin{equation} 
\label{psiDiffequation}
\nabla_\mu (B^{\mu\nu})-\frac{\partial W(\psi_\mu)}{\partial\psi_\nu}=0, 
\end{equation} 
and 
\begin{equation}
\label{phiDiffequation} g^{\mu\nu}\nabla_\mu\nabla_\nu\phi+\frac{\partial V(\phi)}{\partial\phi}=0. 
\end{equation} 
The energy-momentum tensor $T^{\mu\nu}$ satisfies the conservation law: 
\begin{equation} 
\nabla_\nu T^{\mu\nu}=0.
\end{equation}

In the action $S$ we have made the speed of light $c$ a dynamical degree of freedom, in addition to the dynamical degrees
of freedom associated with the metric $g_{\mu\nu}$ and the fields $\psi_\mu$ and $\phi$.

We must now require that local Lorentz invariance and diffeomorphism invariance are violated, so that the speed of light
$c=c(t)$ cannot be made constant by a coordinate transformation as is the case in GR. Let us choose $W(\psi_\mu)$ to be of
the form of a ``Mexican hat'' potential~\cite{Moffat,Moffat2}: 
\begin{equation}
W(\psi_\mu)=-\frac{1}{2}\mu^2\psi_\mu\psi^\mu+\frac{1}{4}\lambda(\psi_\mu\psi^\mu)^2, 
\end{equation} 
where $\lambda > 0$ and $\mu^2 > 0$.  If $W$ has a minimum at 
\begin{equation} 
v_\mu\equiv\psi_\mu=\langle 0\vert\psi_\mu\vert 0\rangle,
\end{equation} 
then the spontaneously broken solution is given by 
\begin{equation}
v^2\equiv\psi^\mu\psi_\mu=\frac{\mu^2}{\lambda}. 
\end{equation} 
We choose the ground state to be described by the timelike vector: 
\begin{equation} 
\psi^{(0)}_\mu=\delta_{\mu 0}v=\delta_{\mu 0}\biggl(\frac{\mu^2}{\lambda}\biggr)^{1/2}.
\end{equation} 
The homogeneous Lorentz group $SO(3,1)$ is broken down to the spatial rotation group $O(3)$. The three
rotation generators $J_i\,(i=1,2,3)$ leave the vacuum invariant, $J_iv_i=0$, while the Lorentz boost generators $K_i$
break the vacuum symmetry $K_iv_i\neq 0$. The spontaneous breaking of the Lorentz and diffeomorphism symmetries produces
massless Nambu-Goldstone modes and massive particle modes~\cite{Bluhm,Kostelecki}. The spontaneous breaking of Lorentz
invariance and diffeomorphism invariance has selected a preferred frame and direction of time.

\section{Initial value conditions and cosmology}

We will work in our application to cosmology in the preferred frame in which the metric is of the FLRW form:
\begin{equation} 
\label{FLRW} 
ds^2=c^2dt^2-a^2\biggl[\frac{dr^2}{1-Kr^2}+r^2(d\theta^2+\sin^2\theta d\phi^2)\biggr],
\end{equation} 
where $K$ is the Gaussian curvature of space and $K=0,+1,-1$ (in units of $[{\rm length}]^{-2}$) for flat,
closed and open models, respectively. The metric has the group symmetry $O(3)\times R$ with a {\it preferred} comoving
time $t$. The energy-momentum tensor $T_{\mu\nu}$ will be described by a perfect fluid: 
\begin{equation}
T^{\mu\nu}=\biggl(\rho+\frac{p}{c^2}\biggr)u^\mu u^\nu-pg^{\mu\nu}, 
\end{equation}
 where $u^\mu=dx^\mu/ds$ is the fluid element four-velocity and $\rho$ and $p$ are the matter density and pressure, respectively.

As in inflationary models, the horizon problem is solved in our VSL model~\cite{Moffat}. Consider a locally flat patch
with the line element: 
\begin{equation} ds^2=c^2dt^2-(dx^i)^2. 
\end{equation} 
From the geodesic equation for light travel
$ds^2=0$, we get 
\begin{equation} dt^2=\frac{1}{c^2}(dx^i)^2. 
\end{equation} 
In the limit when $c\rightarrow\infty$ in an
early phase of the universe, $dt^2\rightarrow 0$ and the Minkowski light cone is squashed. Now all points in an expanding
bubble near the beginning of the universe will be in causal communication with one another. The horizon scale is
determined by 
\begin{equation} 
d_H=ca(t)\int_0^t\frac{dt'}{a(t')}. 
\end{equation} 
Let us assume that for $t > t_c$ we have $c=c_0$. Then for $t < t_c$ we get $d_H\rightarrow \infty$ as $c\rightarrow\infty$ and in the phase $t < t_c$ all points in the expanding spacetime will have been in causal communication.

Setting $\Lambda=0$, the Friedmann equation is given by 
\begin{equation} 
\label{Friedmann} H^2+\frac{Kc^2}{a^2}=\frac{8\pi G\rho}{3}-4H\frac{\dot c}{c}+\frac{8\kappa}{3}\frac{{\dot c}^2}{c^2}, 
\end{equation} 
where $H={\dot a}/{a}$, $\rho=\rho_M+\rho_\psi+\rho_\phi+\rho_r$ and $\rho_r$ denotes the density of radiation.

In standard big-bang cosmology with $c=c_0$: 
\begin{equation} 
\vert\Omega(10^{-43}{\rm sec})-1\vert\sim {\cal O}(10^{-60}),\quad \vert\Omega(1\,{\rm sec})-1\vert\sim {\cal O}(10^{-16}), 
\end{equation} 
where $\Omega=\rho/\rho_c$, $\rho_c=3H^2/8\pi G$ and the Planck time $t_{\rm PL}\sim 10^{-43}\,{\rm sec}$. This implies that the radius of curvature $R_{\rm curv}$ at the Planck time was very large compared to the Hubble radius $R_H=c_0/H$: 
\begin{equation} 
R_{\rm curv}(10^{-43}\,{\rm sec})\sim 10^{30}R_H,\quad R_{\rm curv}(1\, {\rm sec})\sim 10^8R_H. 
\end{equation} 
This means that the universe in standard big-bang cosmology was very special at the Planck time. The universe has survived some $10^{60}$
Planck times without re-collapsing or becoming curvature dominated.

We obtain from (\ref{Friedmann}) the Friedmann equation: 
\begin{equation} 
\label{Omega} \Omega-1=\frac{Kc^2}{{\dot a}^2}+\frac{4}{H}\frac{\dot c}{c}-\frac{8\kappa}{3H^2}\frac{{\dot c}^2}{c^2}, 
\end{equation} 
where $\Omega=8\pi G\rho/3H^2$. In order to resolve the flatness problem, we separate out the
initial value $\Omega_i-1$ and the final value $\Omega_0-1$ in (\ref{Omega}): 
\begin{equation} 
\label{Omegafinal}
\Omega_0-1=\frac{Kc_0^2}{{\dot a}_0^2} 
\end{equation}
 and 
\begin{equation} 
\label{Omegainitial}
\Omega_i-1=\frac{Kc_i^2}{{\dot a}_i^2}+\frac{4}{H_i}\frac{{\dot c}_i}{c_i}-\frac{8\kappa}{3}\frac{{\dot c_i}^2}{H_i^2c_i^2},
\end{equation} 
where $c_0$ is the present speed of light. Dividing (\ref{Omegafinal}) by
(\ref{Omegainitial}) yields 
\begin{equation} 
\label{Flatness} 
\Omega_0=1+\frac{Kc_0^2}{{\dot a}_0^2c^2_i\biggl(\frac{K}{{\dot a}_i^2}+\frac{4{\dot c}_i}{H_ic_i^3}
-\frac{8\kappa{\dot c_i}^2}{3H_i^2c_i^4}\biggr)}|\Omega_i-1|. 
\end{equation}

We have for $c_i=c_0$: 
\begin{equation} 
\Omega_0=1+\frac{{\dot a}_i^2}{{\dot a}_0^2}|\Omega_i-1|. 
\end{equation} 
If ${\dot a}_i/{\dot a}_0 < 10^{-5}$ corresponding to an early phase of inflationary acceleration and
$\vert\Omega_i-1\vert\sim {\cal O}(1)$, then $\Omega_0=1$ to a high accuracy~\cite{Mukhanov}. This solves the flatness
problem in inflationary models\footnote[1]{By implementing a measure Gibbons and Turok~\cite{Turok} find that the
probability of $N_e$ e-folds of inflation is of order $\exp(-3N_e)$.}.

In our VSL model, the flatness problem can also be solved. When in an early phase: 
\begin{equation} 
H_ic_i^3\gg 4{\dot c}_i,\quad 3H_i^2c_i^4\gg 8\kappa {\dot c}_i^2, 
\end{equation}
then we have
\begin{equation}
\Omega_0\sim 1+\frac{c_0^2{\dot a}_i^2}{c_i^2{\dot a}_i^2}|\Omega_i-1|.
\end{equation}
For $c_i\rightarrow\infty$, $\Omega_0=1$ is predicted to a high accuracy. During the radiation dominated phase with $a(t)\propto t^{1/2}$ the radiation density is given by 
\begin{equation}
\label{RadiationDensity} \rho_r=\frac{4\sigma}{c}T^4=\frac{b}{c^3}T^4, 
\end{equation} 
where $\sigma$ is the the Stefan-Boltzmann constant, $b=8\pi^5k_B^4/15 h^3$ and $k_B$ and $h$ are Boltzmann's constant and Planck's constant,
respectively. Thus, $\rho_r\rightarrow 0$ as $c\rightarrow\infty$ and from this we can deduce that $\Omega_i=8\pi
G\rho_r/3H_i^2$ in (\ref{Flatness}) is sufficiently diluted, so that for a large enough initial value of $c_i$ we have
$\Omega_0=1$ to a high degree of accuracy.

In the VSL model, {\it traces of an initial inhomogeneity will be sufficiently smoothed out} as the universe expands.
However, in inflation models bubbles of inflation can be produced that do not inflate sufficiently and these will generate
a non-uniformity problem as the universe expands to the present day~\cite{Steinhardt}.

We have adopted the scenario that in the very early universe the speed of light $c(t)$ has a large value during a short
time duration when $t_i < t < t_c$, and it has the value $c=c_0$ for $t > t_c$ when the Friedmann equations and the
cosmology are described by GR.

Let us assume for our scalar fluctuation ``seed'' field $\phi$ that in (\ref{phiDiffequation}) $V(\phi)=0$, yielding
\begin{equation} 
\ddot\phi+3H\dot\phi=0. 
\end{equation} 
The solution to this equation is 
\begin{equation} 
\label{dotphi}
\dot\phi=\sqrt{12}B\biggl(\frac{a_*}{a}\biggr)^3, 
\end{equation} 
where $B$ is a constant and $a_*$ is a reference value
for $a$. For large $\dot\phi$ the kinetic contribution to $\rho_\phi\propto\frac{1}{2}(\dot\phi)^2 $ will dominate the
matter densities $\rho_M$ and $\rho_\psi$ and the Friedmann equation (\ref{Friedmann}) for $t > t_c$ and $c=c_0$ becomes
\begin{equation} H^2+\frac{Kc_0^2}{a^2}=\frac{1}{12}\dot\phi^2. 
\end{equation} 
Substituting the solution (\ref{dotphi}),
we get \begin{equation} H^2+\frac{Kc_0^2}{a^2}=B^2\biggl(\frac{a_*}{a}\biggr)^6. 
\end{equation} 
Because the field $\phi$ dominates in the early universe, we can neglect the spatial curvature for $c=c_0$ and we obtain the approximate solution
given by 
\begin{equation} 
\label{asolution} a(t)=a_*(3Bt)^{1/3},\quad H(t)=\frac{1}{3t}. 
\end{equation} 
The equation of state for a massless scalar field gives for the exponent $n$ in $a(t)\propto t^n$ the value $n=2/3(1+w)$, so for $n=1/3$
we get $w=1$. 

We observe that there is no explicit source for the field $\phi$ in Eq.(\ref{phiDiffequation}), which implies that as the
universe expands the field $\phi$ becomes increasingly diluted to the point where it has unobservable effects at present.
By performing a post-Newtonian expansion of the gravitational field in the solar system, we have: $g_{00}\approx 1
+2GM_\odot/c_0^21\,{\rm AU}$. If the contribution of the field $\phi$ is to be significant in the solar system, then
$2GM_\odot/c_0^21\,{\rm AU}\propto\dot\phi^2$. However, as the universe expands we have according to (\ref{dotphi}) that
$\dot\phi\sim 1/a^3$, so that the effects of the scalar field $\phi$ will become unobservable in the present universe.

\section{Primordial fluctuations and gravitational waves}

Let us consider in our VSL model the mechanism for the generation of cosmological fluctuations and the growth of large
structures without inflation. In inflationary models, scalar quantum fluctuations oscillate until their wavelengths become
equal to the Hubble radius $R_H=c_0/H$. When they pass beyond $R_H$ the oscillations are damped and the fluctuation modes
are ``frozen'' as classical fluctuations with amplitude $\delta\phi\sim H/2\pi$~\cite{Mukhanov}. In an inflationary
spacetime the wavelengths of quantum field fluctuations $\delta\phi$ are stretched by rapid expansion: \begin{equation}
\lambda_f\propto a(t)\propto\exp(Ht), \end{equation} where $H$ is approximately constant. Short-wavelength fluctuations
are quickly redshifted by the inflationary expansion until their wavelengths are larger than the size of the horizon
$R_H$.

In our VSL model, the wavelengths of the quantum field fluctuations $\delta\phi$ are stretched by the short duration large
increase of $c$: \begin{equation} \lambda_f=\frac{c}{\nu_f}, \end{equation} where $\nu_f$ is the frequency. The redshifted
wavelengths of the short-wavelength fluctuations become larger than the horizon and are frozen in as classical
fluctuations. The amplitudes of quantum modes are calculated at the horizon crossing, when the wavelength of a mode is
equal to $a/k=c_0/H$. For an inflationary epoch with 60 e-folds of inflation of the cosmic scale $a$, the initial matter
fluctuations and gravitational waves with wavelengths $\lambda_i$ are stretched to their wavelengths today by the amount:
\begin{equation} \lambda_0\sim\exp({\cal N})\lambda_i, \end{equation} where ${\cal N}$ is the number of e-folds of
inflation, ${\cal N}\geq 60$. In the VSL model, the initial fluctuation wavelengths $\lambda_i$ are stretched by an
equivalent amount: \begin{equation} \lambda_0\sim{\cal Q}\lambda_i, \end{equation} where ${\cal Q}\gtrsim 10^{30}$.

The fluctuations are of two kinds: scalar matter quantum fluctuations and tensor gravitational wave fluctuations.  Let us
consider first the scalar fluctuations. We will consider a scenario in which we assume that $V(\phi)=0$. The fluctuations
$\delta\phi(t,\vec x)$ about a cosmological background will be considered in the comoving frame with the minimally-coupled
Klein-Gordon field equation: 
\begin{equation} 
\label{phipert} 
\frac{d^2\delta\phi_{\vec{k}}}{dt^2}+3H\frac{d
\delta\phi_{\vec{k}}}{dt} +\frac{c_0^2k^2}{a^2}\delta\phi_{\vec{k}}=0, 
\end{equation} 
where $k=|{\vec k}|$ and
\begin{equation} 
\label{Fourier} \delta\phi(\vec x) = (2\pi)^{-2/3}\int d^3{\vec k}\,\exp[-i\vec{k}\cdot\vec{x}]\delta\phi_{\vec{k}}. 
\end{equation}

By substituting the solution (\ref{asolution}) into (\ref{phipert}), we get the equation of motion: 
\begin{equation}
\label{a(t)equation} 
\frac{d^2\delta\phi_{\vec{k}}}{dt^2}+\frac{1}{t}\frac{d \delta\phi_{\vec{k}}}{dt}
+\frac{c_0^2k^2}{a_*^2(3Bt)^{2/3}}\delta\phi_{\vec{k}}=0. 
\end{equation} 
This equation has the general Bessel function
solution~\cite{Moffat5}: 
\begin{equation} 
\label{Besselsolution} 
\delta\phi_{\vec{k}}(y) = A_1J_0(y)+A_2Y_0(y),
\end{equation} 
where $A_1$ and $A_2$ are constant coefficients and \begin{equation} \label{yvalue} y=\frac{c_0
k}{2a_*H_\ell}\biggl(\frac{a}{a_*}\biggr)^2. \end{equation} Here, $H_\ell=(3B)^{2/3}/3a_*$ and we define
$H_\ell=c_0/\ell_0$, where $\ell_0$ is the length scale at which the normalized wave function is in its ground state.

We can obtain from (\ref{Besselsolution}) the normalized plane wave solution: 
\begin{equation} 
\label{classicalwave}
\delta\phi_{\vec{k}}(t_k) = \biggl(\frac{8\pi G\hbar}{c_0^2(2\pi a_k)^3\omega_k}\biggr)^{1/2} \cos(\omega_k t_k
-\vec{k}\cdot\vec{x} +\delta), 
\end{equation} 
where 
\begin{equation} 
\label{omegaeq}
\omega_k =\frac{c_0 k}{a_k}. 
\end{equation} 
This solution is approximately equivalent to the one we get from adopting a flat
Minkowski spacetime. We assume that after the fluctuation modes cross the horizon they are in a classical state in the
Minkowski spacetime\footnote[2]{Hollands and Wald~\cite{Hollands} considered a model which did not have a VSL or
inflationary mechanism to stretch the quantum fluctuation wavelengths to super-horizon lengths. They assumed that the
classical fluctuations were born super-horizon, were frozen at a length scale $\ell_0$ and were described by a plane wave
in its ground state. Due to the lack of a VSL or inflationary dilution of the radiation density, the value of the
radiation density at the epoch when the fluctuations were born was unphysically large~\cite{Kofman}. As demonstrated in
Eq.(\ref{RadiationDensity}), the VSL mechanism as in the case of inflation dilutes the radiation density.}.

The scale at which the fluctuation mode exists is given by the condition: 
\begin{equation} 
\label{Condition} 
a_k=k\ell_0.
\end{equation} 
From this condition, we get from (\ref{yvalue}): 
\begin{equation} 
\label{yk} 
y_k=\gamma k^3, 
\end{equation}
where $y_k$ is the value of $y$ evaluated for $a=a_k$ and $\gamma=\ell_0^3/2a_*^3$.

If we assume that we should use (\ref{classicalwave}) as initial data for the classical solution of (\ref{phipert}), then
we will match not only the initial perturbation, but also its time derivative. Keeping only the dominant contribution as
$y\rightarrow 0$ and $\omega_k=c_0k/a_k$ gives 
\begin{equation} 
\label{Matchedsolution} 
\delta\phi_{\vec{k}}\approx
\biggl(\frac{9\pi G\hbar}{2c_0^2(2\pi a_k)^{3}\omega_k}\biggr)^{1/2} \cos(\omega_k t_k -\vec{k}\cdot\vec{x}
+\delta)\ln(y_k)J_0(y_k). 
\end{equation} 
From (\ref{omegaeq}), (\ref{yk}) and the Planck length, 
$\ell_{\rm PL}=(G\hbar/c_0^3)^{1/2}\sim 10^{-33}\,{\rm cm}$, we obtain the density perturbation power spectrum: 
\begin{equation}
\label{Powerspectrum} 
{\cal P}_{\delta\phi}=\frac{9}{2(2\pi)^3}\left(\frac{\ell_{\rm PL}^2}{\ell_0^2}\right)\ln^2(y_k).
\end{equation} 
Recalling that $\dot\phi^2\sim 12 H^2$, we obtain the curvature power spectrum:
\begin{equation}
\label{curvpowerspectrum} 
{\cal P}_{\cal R}(k)=\biggl(\frac{H^2}{\dot\phi^2}\biggr){\cal P}_{\delta\phi}=\frac{3}{8(2\pi)^3}\left(\frac{\ell_{\rm PL}^2}{\ell_0^2}\right)\ln^2(y_k). 
\end{equation} 
Moreover, we have 
\begin{equation} {\cal P}_{\cal R}(k)=2\pi^2A_s\biggl(\frac{k}{k_*}\biggr)^{n_s-1}, 
\end{equation} 
where $A_s$ is the scalar fluctuation amplitude and $n_s$ is the spectral index. The power spectrum (\ref{curvpowerspectrum}) is scale
invariant except for the factor $\ln^2(y_k)$. The non-scale invariant contribution results from matching the initial state
and its time derivative, and that the Bessel function $Y_0(y_k)$ is logarithmically divergent when $y_k\rightarrow 0$.
This will lead to a slight deviation from a scale invariant spectrum.

The scalar mode spectral index is given by 
\begin{equation} 
\label{spectral index} n_s=1+\frac{d\ln\mathcal{P}_{\cal
R}}{d\ln k}. 
\end{equation} 
From (\ref{yk}) and (\ref{curvpowerspectrum}), we get 
\begin{equation} 
\label{Spectralresult}
n_s=1+\frac{6}{\ln(y_k)}. 
\end{equation} 
The running of the spectral index is calculated from 
\begin{equation}
\alpha_s=dn_s/d\ln k, 
\end{equation} 
which yields 
\begin{equation} 
\alpha_s=-\tfrac{1}{2}(1-n_s)^2. 
\end{equation}

We have in the large scale limit the anisotropic amplitude: 
\begin{equation} 
\label{eq:delta H} \delta_H=
\frac{2}{5}\sqrt{\mathcal{P}_{\cal R}} \approx\frac{2}{5}\biggl(\frac{3}{8(2\pi)^3}\biggr)^{1/2}\biggl(\frac{\ell_{\rm PL}}{\ell_0}\biggr)|\ln(y_k)|. 
\end{equation} 
By adopting 
\begin{equation}
\frac{2}{5}\biggl(\frac{3}{8(2\pi)^3}\biggr)^{1/2}|\ln(y_k)|\sim {\cal O}(1), 
\end{equation} 
we have $\delta_H\sim\ell_{\rm PL}/\ell_0$. Fixing the length scale ${\ell }$ to be $\ell_0\sim 10^5\ell_{\rm PL}\sim 10^{-28}\,{\rm cm}$,
which is of the order of the grand unification scale, we can match the amplitude of the observed CMB fluctuations,
$\delta_H\sim 10^{-5}$.

For the value $\ln(y_k)\sim -150$, we obtain the result for the spectral index:
\begin{equation}
n_s\sim 0.96. 
\end{equation} 
This is in good agreement with the result obtained by Planck
Mission~\cite{Planck2015}: 
\begin{equation} n_s=0.9603\pm 0.0073. 
\end{equation} 
For the running of the spectral index $n_s$, we get 
\begin{equation} 
\alpha_s\sim - 8\times 10^{-4}, 
\end{equation} 
which is in approximate agreement with the Planck Mission result: 
\begin{equation} 
\alpha_s=-0.013\pm 0.0090. 
\end{equation}

In inflationary models the derivation of the power spectrum and the spectral index depend sensitively on the shape of the
inflaton potential $V(\phi_{\rm inflaton})$ and its derivatives with respect to $\phi$. The condition of a slow-roll
potential is required to produce enough e-folds of inflation. This is not the case in our VSL derivation of the power
spectrum and the spectral index. Our derivation of the scalar fluctuation power spectrum does not depend sensitively on
the shape of the potential $V(\phi)$. This can reduce the VSL model dependence and associated fine-tuning problems.

We now turn to the spectrum of relic gravitational waves. The conformally flat background metric is 
\begin{equation}
ds^2=a^2(\eta)(c^2d\eta^2-(dx^i)^2). 
\end{equation} 
In our model the wavelength of gravity waves is given by
$\lambda_g=c_g/\nu_g$, where $\nu_g$ is the frequency of a gravitational wave. During the phase when $c_g\gg c_0$
gravitational waves are generated and their wavelengths are stretched and cross the horizon. As the universe expands, the
amplitude of the gravitational wave spectrum passes back into the observable universe and we can observe a B-polarization
with a non-zero spectral index $n_t$ and ratio $r$=tensor/scalar.

The tensor perturbations can be expressed as 
\begin{equation} 
ds^2=a^2[c^2d\eta^2-(\delta_{ij}+2h_{ij})dx^idx^j],
\end{equation} 
where 
\begin{equation} h_{ij}=\int\frac{d^3{\vec
k}}{(2\pi)^{3/2}}\sum^2_{\lambda=1}\psi_{k,\lambda}e_{ij}({\vec k,\lambda}\exp(i{\vec k}\cdot{\vec x})), 
\end{equation}
and where $e_{ij}({\vec k},\lambda)$ is a polarization tensor satisfying $e_{ij}=e_{ji}, e_{ii}=0, k_ie_{ij}=0$ and
$e_{ij}({\vec k},\lambda)e^*_{ij}({\vec k},\mu)=\delta_{\lambda\mu}$. Moreover, we have 
\begin{equation}
\langle\psi_{{\vec k},\lambda},\psi^*_{{\vec l},\lambda}\rangle=2\pi^2{\cal P}_{\delta\psi}\delta({\vec k}-{\vec l}),
\end{equation} 
where ${\cal P}_{\delta\psi}$ is the gravitational wave spectrum.

The gravitational tensor component can be expressed as the superposition of two scalar polarization wave modes:
\begin{equation} 
h^{\rm tensor}_{+,\times}=\phi_{+,\times} = {\rm scalar}, 
\end{equation} 
where $+,\times$ refer to the longitudinal and transverse polarization modes, respectively. The modes obey the scalar equation of motion at the horizon
and super-horizon scales when $c=c_0$: 
\begin{equation} 
\label{phigrav}
\frac{d^2\phi_{+,\times,\vec{k}}}{dt^2}+3H\frac{d\phi_{+,\times,\vec{k}}}{dt}
+\frac{c_0^2k^2}{a^2}\phi_{+,\times,\vec{k}}=0. 
\end{equation} 
The sizes of the second and third terms depend on the
magnitude of $c_0 k/aH$. If $c_0 k/aH\rightarrow 0$, then the gravitational wave mode is outside the horizon and we can
neglect the third term and the solution to the wave equation approaches a constant. If $c_0 k/aH$ is large, the mode is
inside the horizon and the second term becomes sub-leading. The mode then undergoes a damped oscillation and decays as
$1/a$. The tensor modes of interest are outside the horizon with constant values determined by the primordial
distribution, generated at a sub-horizon scale during the phase when $c_g$ has a large value $c_g\gg c_0$. During the
radiation and matter domination epochs the modes gradually re-enter the horizon and damp away. Only the tensor modes which
entered the horizon just before the surface of last scattering lead to important effects in the CMB.

The tensor mode power spectrum is \begin{equation} {\cal P}_t(k)={\cal P}_\psi(k_p)\biggl(\frac{k}{k_p}\biggr)^{n_t},
\end{equation} where $k_p=0.004\,{\rm Mpc}^{-1}$ is the pivotal scale. By solving the scalar wave equation (\ref{phigrav})
in the same manner as was done for the scalar matter fluctuations, we obtain the power spectrum: 
\begin{equation} 
{\cal P}_t(k)=k^3{\cal P}_{\phi_{+,\times}}(k) =\frac{N}{(2\pi)^3}\left(\frac{\ell_{\rm PL}^2}{\ell_0^2}\right)\ln^2(y_k),
\end{equation} 
where $N$ is a constant.  As in the case of the scalar matter fluctuations, the gravitational wave mode
fluctuations are scale invariant up to the slight scale breaking factor $\ln^2(y_k)$.

The tensor mode spectral index is given by 
\begin{equation} 
\label{Gravspectra} 
n_t=\frac{d\ln{\cal P}_t}{d\ln k}.
\end{equation} 
This yields the result 
\begin{equation} 
n_t=\frac{6}{\ln (y_k)}. 
\end{equation} 
Choosing, as before for the scalar perturbations, the value $\ln (y_k)\sim -150$ we obtain 
\begin{equation} 
\label{Tensorindex} n_t=-0.04.
\end{equation} 
This spectral index (tilt) $n_t$ is red which agrees with the standard inflationary model result for $n_t$. 
A measurement of the tensor mode tilt $n_t$ is needed to give a definitive test of the models. A joint analysis of the BICEP2/Keck Array observational data found strong evidence for foregound dust and no statistical significant evidence for gravitational wave tensor modes. An upper limit was obtained for the tensor-to-scalar mode ratio $r < 0.12$~\cite{BICEP2}.

If we now adopt the result $r < 0.12$~\cite{BICEP2}, then we get $|r/n_t| < 3$. The single-field inflationary model
determines a consistency condition by the slow-roll parameter, $\epsilon=-\dot H/H^2$, related to the equation of state
$p=\epsilon-\rho$. For $r=16\epsilon$ and $n_t=-2\epsilon$ the model gives $|r/n_t|=8$, which is satisfied irrespective of the
form of the single-field inflationary potential.

\section{Non-gaussianities in the VSL cosmology}

In this section, we first briefly review the standard process of calculating nonlinear perturbations by virtue of the so-called $\delta{\cal N}$ formalism.
Within the context of the VSL cosmology, primordial fluctuations are seeded by the scalar field $\phi$ and, therefore, the mechanism of generating
primordial curvature perturbation in the VSL cosmology is analogues to the curvaton scenario~\cite{Lyth:2001nq, Mollerach:1989hu, Linde:1996gt,
Enqvist:2001zp}. Then, with the assumption that the end of the epoch $c\gg c_0$ when $c=c_0$ is a uniform total density slice, we generalize the
$\delta{\cal N}$ formalism~\cite{Salopek:1990jq, Sasaki:1995aw, Sasaki:1998ug, Wands:2000dp, Lyth:2004gb} to be applicable to our case. Afterwards, we
analyze the non-gaussianities of primordial curvature perturbations in VSL cosmology.

\subsection{The generalized $\delta{\cal N}$ Formalism}

The $\delta {\cal N}$ formalism is available upon two conditions: firstly, the universe is isotropic and homogenous at extremely large scales that can accommodate a large number of causally isolated regions; secondly, the perturbation is frozen after the horizon exit. These two conditions are easily met in inflationary cosmology. However, for alternatives to inflation, if one can argue based on them, the $\delta {\cal N}$ formalism is still available. The first condition is in fact related to the horizon problem and, hence can be satisfied in quite a number of alternative models. For the second condition, it is not satisfied in matter bounce cosmology and marginally satisfied in ekpyrotic cosmology. In VSL models, the universe is still expanding with an effective equation of state parameter $w=1$, and as a result, the dominant perturbation grows as a logarithmic function. In this case, the second condition is also satisfied and from this argument the $\delta {\cal N}$ formalism is available as well in VSL. 

On superhorizon scales, if we assume negligible interaction between primordial scalar field $\phi$ and other matter fields, the component curvature
perturbation on the uniform density slice can be identified as 
\begin{eqnarray} 
\label{delta_N}
 \zeta_i (x) = \delta{\cal N} (x) + \frac{1}{3} \int_{\bar\rho(t)}^{\rho_i(x)} \frac{d \tilde\rho_i}{\tilde\rho_i+P_i(\tilde\rho_i)},
\end{eqnarray} 
through the $\delta{\cal N}$ formalism, where the subscript $i$ represents the component matter field in the universe.

Having the component curvature perturbation in mind, we can calculate the curvature perturbation on the uniform density slice. In analogy with the curvaton mechanism, it is natural to choose this slice at the end moment of $c\gg c_0$ period when $c=c_0$. On the uniform density slice we have \begin{eqnarray}
 \rho_r+\rho_\phi = \bar\rho,
\end{eqnarray} 
where $\rho_r$ and $\rho_\phi$ represent the radiation and scalar field densities, respectively. For the primordial scalar field, the
background dynamics behaves as a stiff fluid with an effective equation of state $w_\phi = 1$. Then, following Eq. \eqref{delta_N}, we can
obtain~\cite{Cai:2010rt}: 
\begin{eqnarray}
 \rho_r = \bar\rho_r e^{4(\zeta_r-\zeta)},\quad\rho_\phi = \bar\rho_\phi e^{6(\zeta_\phi-\zeta)}.
\end{eqnarray} 
As a consequence, the curvature perturbation $\zeta$ can be derived on the uniform density slice through the relation: 
\begin{eqnarray}
\zeta = \tilde{r} \zeta_\phi , 
\end{eqnarray} 
where \begin{eqnarray} \tilde{r} = \frac{3\Omega_\phi}{2+\Omega_\phi}. 
\end{eqnarray} 
Here, we have introduced the transfer efficiency coefficient $\tilde{r}$ and $\Omega_\phi \equiv \bar\rho_\phi/\bar\rho_{tot}$ is the density parameter for the primordial scalar field. Note that, in a generic case, $\tilde{r} = 3(1+w_\phi)\Omega_\phi/[4+(3w_\phi-1)\Omega_\phi]$, and we have applied the relation $w_\phi =1$. In the VSL cosmology, we have $\Omega_\phi = 1$, for the scalar field dominates over other matter fields at the end of the primordial era. This implies that the transfer from the scalar field fluctuations into primordial curvature perturbations is very efficient and thus, $\tilde{r} = 1$.

Within the local ansatz of the curvature perturbation, one can expand its form order by order as follows, 
\begin{align}
 \zeta(x) &= \zeta_1(x) + \zeta_2(x) + \zeta_3(x) + O(\zeta_4) \nonumber\\
          &= \zeta_1(x) +\frac{3}{5}f_{\rm NL}\zeta_1^2(x) +\frac{9}{25}g_{\rm NL}\zeta_1^3(x) + O(\zeta_4),
\end{align} 
where $\zeta_1$ is the fluctuation of the Gaussian distribution, and $\zeta_n$ for $n > 1$ are the non-Gaussian fluctuations. From the above expansion, we
obtain 
\begin{eqnarray}
\label{fg_NL}
 f_{\rm NL} = \frac{5}{3}\frac{\zeta_2}{\zeta_1^2} , \quad g_{\rm NL} = \frac{25}{9}\frac{\zeta_3}{\zeta_1^3},
\end{eqnarray} 
for the local type. To be general, one can relate the nonlinearity parameters to the bispectrum and the trispectrum via: \begin{align}
 &B({\bf k}_1, {\bf k}_2, {\bf k}_3) = \frac{6}{5} f_{\rm NL} [P(k_1)P(k_2)+2 {\rm perm}]~, \nonumber\\
 &T({\bf k}_1, {\bf k}_2, {\bf k}_3, {\bf k}_4) = \frac{54}{25} g_{\rm NL} [P(k_1)P(k_2)P(k_3)+3 {\rm perm}] \nonumber\\
 & ~~~~~~~~~~ +\tau_{\rm NL}[P(k_1)P(k_2)P(|{\bf k}_1+{\bf k}_3|) +11 {\rm perm}],
\end{align} 
where these spectra are associated with the correlation functions as follows: 
\begin{align}
 &\langle \zeta({\bf k}_1)\zeta({\bf k}_2) \rangle = (2\pi)^3 P(k_1) \delta^{(3)}(\sum_{a=1}^2{\bf k}_a) ~,\\
 &\langle \zeta({\bf k}_1)\zeta({\bf k}_2)\zeta({\bf k}_3) \rangle = (2\pi)^3 B({\bf k}_1, {\bf k}_2, {\bf k}_3) \delta^{(3)}(\sum_{a=1}^3{\bf k}_a) ~,
 \nonumber\\
 &\langle \zeta({\bf k}_1)\zeta({\bf k}_2)\zeta({\bf k}_3)\zeta({\bf k}_4) \rangle\nonumber\\
&~~~~~~~~~= (2\pi)^3 T({\bf k}_1, {\bf k}_2, {\bf k}_3, {\bf k}_4) \delta^{(3)}(\sum_{a=1}^4{\bf k}_a).\nonumber \end{align} 
Moreover, the two point correlation function is related to the regular power spectrum through the relation: 
\begin{eqnarray}
 {\cal P}_\zeta (k) = \frac{k^3}{2\pi^2} P(k).
\end{eqnarray}

\subsection{Primordial perturbations and non-Gaussianities in the VSL cosmology}

Following the discussion in the previous subsection, one can find that the dynamics of primordial curvature perturbation would be identified by the
variation of the generalized VSL e-folding number by making use of the local ansatz. To calculate such a variation, we need to know the evolution of the
background universe. The scalar field satisfies the Klein-Gordon equation: 
\begin{eqnarray}\label{KG_fast}
 \ddot\phi +3 H\dot\phi = 0.
\end{eqnarray} 
In this phase, $a\propto t^{1/3}$ and, therefore, the Hubble parameter is expressed as 
\begin{eqnarray} 
\label{H_fast}
 H(t) = \frac{1}{3t}.
\end{eqnarray}

Substituting Eq. \eqref{H_fast} into the background equation of motion \eqref{KG_fast} yields the following solution: 
\begin{eqnarray} 
\label{phigamma} 
\phi(t) = \phi_f + \gamma \ln\bigg(\frac{t}{t_f}\bigg),\quad \dot\phi (t) = \frac{\gamma}{t}, 
\end{eqnarray} 
where $\phi_f$ denotes the value of the scalar field at the end moment of the
phase $t_f$ when $c=c_0$. The coefficient $\gamma$ is an integration constant that can be determined by the background Friedmann equation\footnote[3]{The constant $\gamma$ in (\ref{phigamma}) is not to be confused with the $\gamma$ in (\ref{yk})}. Without loss of generality, we keep $\gamma$ as a free coefficient and see its effect on the variation of the effective e-folding number.

In our case, the VSL equivalent of the e-folding number is given by 
\begin{eqnarray}
 {\cal N} \equiv \int_t^{t_f} H dt = \frac{1}{3} \ln \bigg( \frac{t}{t_f} \bigg),
\end{eqnarray} 
and, correspondingly, the trajectories of the background dynamics can either be described by the phase space of $(\phi, \dot\phi)$ or that
of $({\cal N}, \gamma)$. In this regard, the method of calculating $\delta{\cal N}$ is similar to the analysis of~\cite{Chen:2013eea}.  However, the
effects of the background evolution of the VSL cosmology upon curvature perturbations are analogous to those occurring in the study of the matter bounce cosmology~\cite{Cai:2009fn, Cai:2011zx}, {\it where there is no quasi de Sitter expansion}. To be explicit, from now on we switch to the phase space of $(\phi, \gamma)$. Consequently, the VSL equivalent of the e-folding number takes the form: 
\begin{eqnarray}
\label{N_efold}
{\cal N} = {\cal N}(\phi, \lambda) = \frac{\phi-\phi_f}{3\gamma}.
\end{eqnarray}

By expanding the scalar field $\phi$, and determining the integration constant $\gamma$ via $\phi\rightarrow \phi+\delta\phi$ and $\gamma\rightarrow
\gamma+\delta\gamma$, we can determine the curvature perturbations order by order up to third order: 
\begin{align}
\label{zeta_123}
 &\zeta_1 = {\cal N}_{,\phi}\delta\phi, \\
 &\zeta_2 = \frac{1}{2}{\cal N}_{,\phi\phi}\delta\phi^2 + {\cal N}_{,\phi\gamma}\delta\phi\gamma + \frac{1}{2}{\cal N}_{,\gamma\gamma}\delta\gamma^2,
 \nonumber\\
 &\zeta_3 = \frac{1}{6}{\cal N}_{,\phi\phi\phi}\delta\phi^3 + \frac{1}{3}{\cal N}_{,\phi\phi\gamma}\delta\phi^2\gamma
+ \frac{1}{3}{\cal N}_{,\phi\gamma\gamma}\delta\phi\gamma^2 \frac{1}{6}{\cal N}_{,\gamma\gamma\gamma}\delta\gamma^3.\nonumber 
\end{align} 
In the above expression, the subscript $_{,\phi}\equiv \partial/\partial\phi$ denotes the derivative with respect to $\phi$. After having obtained the expression ${\cal N}$ in \eqref{N_efold}, we can  explicitly determine the coefficients such as ${\cal N}_{,\phi}$. The above expression automatically includes the assumption that the field fluctuations are described before the primordial era by a highly Gaussian distribution.

From \eqref{phigamma} we get 
\begin{eqnarray}
 \gamma = \gamma(\phi, \dot\phi) = \frac{\phi-\phi_f}{{\cal W}\bigg(\frac{\phi-\phi_f}{t_f\dot\phi}\bigg)},
\end{eqnarray} 
where ${\cal W}$ is the Lambert function. As a consequence, we can find \begin{eqnarray}
 \delta\gamma \simeq \gamma_{,\phi}\delta\phi = \frac{\gamma}{\gamma+\phi-\phi_f}\delta\phi,
\end{eqnarray} where the contribution of $\delta\dot\phi$ is secondary at superhorizon scales. Inserting the above expression into Eq. \eqref{zeta_123}, we can derive 
\begin{align}
 &\zeta_1 = -\frac{1}{3(\gamma+\phi-\phi_f)} \delta\phi ~, \\
 &\zeta_2 = \frac{1}{3(\gamma+\phi-\phi_f)^2} \delta\phi^2 ~, \\
 &\zeta_3 = -\frac{(2\gamma-\phi+\phi_f)}{9\gamma(\gamma+\phi-\phi_f)^3} \delta\phi^3.
\end{align}

Recall that the nonlinearity parameters satisfy the relations in \eqref{fg_NL}. Consequently, these parameters can be identified in the VSL cosmology as
follows: 
\begin{align} \label{fg_NL_simple}
 &f_{\rm NL} = 5, \nonumber\\
 &g_{\rm NL} = \frac{25(2\gamma-\phi+\phi_f)}{3\gamma}.
\end{align} 
From Eq. \eqref{fg_NL_simple}, we observe that the nonlinearity parameter at second order $f_{\rm NL}$ is a positive constant of order unity
and thus is almost scale invariant.  However, the nonlinearity parameter at third order $g_{\rm NL}$ is a function of the scalar field during the
primordial phase and therefore implies a strong scale dependence.

\section{Conclusions}

We have formulated a VSL model in which the homogeneous Lorentz group $SO(3,1)$ is spontaneously broken to the rotation
group $O(3)$ by the non-zero vacuum expectation value $\langle 0\vert\psi_\mu\vert 0\rangle$. This determines a preferred
time $t$ in the cosmological model corresponding to the comoving time in an FLRW spacetime. 

In contrast to the inflationary scenario, our VSL model prediction of the almost scale invariant, Gaussian fluctuation
spectra for matter and relic gravitational waves does not rely on determining the shape of a potential and its
derivatives. The model can relieve the fine-tuning that is inevitably a consequence of inflationary models. Once the short
duration phase of $c\gg c_0$ and $c_g\gg c_{g0}$ has taken place and the wavelengths of the initial primordial quantum
fluctuations and gravitational waves have been stretched through the horizon, then classical solutions of the wave equations 
for the quantum and gravitational wave fluctuations can be employed to generate the power spectra and the spectral indices 
of the scalar and tensor fluctuation modes. Our VSL model can produce the non-zero scalar and tensor spectral indices $n_s=0.96$ 
and $n_t=-0.04$. Combined with the bound on the $r=tensor/scalar$ ratio, $r < 0.12$, we obtain $|r/n_t| < 3$.

We have derived in our variable speed of light (VSL) model the primordial nonlinear fluctuations. We find that the parameter $f_{NL}=5$ 
which is compatible with the Planck2015 result: $f_{NL}=2.7\pm 5.8$~\cite{PLANCK2015}. The parameter $g_{NL}$ evolves during the 
primordial era and displays a running behavior. These results will allow the VSL model to be distinguished from inflationary models and other 
models such as the bouncing and string gas cosmology models.

Since the ordered phase in the spontaneous symmetry breaking of Lorentz invariance is at a much lower entropy than the
restored, disordered symmetry phase and due to the existence of a domain determined by the direction of the vev, $\langle
0\vert\psi_\mu\vert 0 \rangle$, a natural explanation is given for the cosmological arrow of time and the origin of the
second law of thermodynamics~\cite{Moffat,Moffat2,Moffat10}. The ordered state of low entropy in the symmetry broken phase
with $c\gg c_0$, becomes a state of high entropy in the symmetry restored disordered phase with $c = c_0$. The spontaneous
symmetry breaking of the gravitational vacuum leads to a manifold with the structure $O(3)\times R$, in which time appears
as an absolute external time parameter. The vev $\langle 0\vert\psi_\mu\vert 0 \rangle$ points in a chosen direction of
time to break the symmetry of the vacuum creating an arrow of time.

\section*{Acknowledgments}

I thank Andrew Liddle, Jerome Martin, Michael Clayton, Yi-Fu Cai, Robert Brandenburger, Viktor Toth and Martin Green for helpful discussions. This research
was generously supported by the John Templeton Foundation. Research at the Perimeter Institute for Theoretical Physics is
supported by the Government of Canada through industry Canada and by the Province of Ontario through the Ministry of
Research and Innovation (MRI). I thank the Institut d'Astrophysique de Paris (IAP) where part of this research was completed for
their hospitality.

\end{document}